\documentclass[twocolumn]{aastex62_tweeked_for_nat}

\usepackage[sort&compress]{natbib}

\usepackage[utf8]{inputenc}
\usepackage{graphicx}
\usepackage{amssymb,amsmath}
\usepackage{upgreek} 
\usepackage{xcolor}
\usepackage{soul}
\usepackage{hyperref}
\usepackage{fancyvrb}
\usepackage[normalem]{ulem}
\usepackage[title]{appendix}
\usepackage{lineno}

\begin{document}
\title{High-energy neutrinos from the vicinity of the supermassive black hole in NGC 1068}

\author[0000-0002-4707-6841]{P. Padovani} \affiliation{European Southern Observatory, Karl-Schwarzschild-Stra{\ss}e 2, D-85748 Garching bei M{\"u}nchen, Germany}
\author[0000-0003-0705-2770]{E. Resconi}
\affiliation{Technical University of Munich, TUM School of Natural Sciences, Department of Physics, James-Franck-Stra{\ss}e 1, D-85748 Garching bei M{\"u}nchen, Germany}
\author[0000-0002-6584-1703]{M. Ajello}
\affiliation{Department of Physics and Astronomy, Clemson University, Clemson, SC 29631, USA}
\author[0000-0001-8525-7515]{C. Bellenghi}
\affiliation{Technical University of Munich, TUM School of Natural Sciences, Department of Physics, James-Franck-Stra{\ss}e 1, D-85748 Garching bei M{\"u}nchen, Germany}
\author[0000-0002-4622-4240]{S. Bianchi}
\affiliation{Dipartimento di Matematica e Fisica, Universit\`a degli Studi Roma Tre, Via della Vasca Navale 84, I-00146, Roma, Italy}
\author[0000-0003-2480-599X]{P. Blasi}
\affiliation{Gran Sasso Science Institute, Viale F. Crispi 7, I-67100, L'Aquila, Italy}
\affiliation{INFN-Laboratori Nazionali del Gran Sasso, Via G. Acitelli 22, Assergi (AQ), Italy}
\author[0000-0002-1227-8435]{K.-Y. Huang}
\affiliation{Leiden Observatory, Leiden University, P.O. Box 9513, 2300 RA Leiden, The Netherlands}
\author{S. Gabici}
\affiliation{Université Paris Cité, CNRS, Astroparticule et Cosmologie, 10 Rue Alice Domon et Léonie Duquet, F-75013 Paris, France}
\author{V. G\'{a}mez Rosas}
\affiliation{Sterrewacht Leiden, Niels Bohrweg 2, 2333 CA Leiden, The Netherlands}
\author[0000-0002-9566-4904]{H. Niederhausen}
\affiliation{Department of Physics and Astronomy, Michigan State University, East Lansing, MI 48824, USA}
\author[0000-0003-0543-0467]{E. Peretti}
\affiliation{Niels Bohr International Academy, Niels Bohr Institute,University of Copenhagen, Blegdamsvej 17, DK-2100 Copenhagen, Denmark}
\affiliation{Université Paris Cité, CNRS, Astroparticule et Cosmologie, 10 Rue Alice Domon et Léonie Duquet, F-75013 Paris, France}
\author[0009-0000-3061-1118]{B. Eichmann}
 \affiliation{Ruhr-Universit\"at Bochum, Theoretische Physik IV, Fakult\"at f\"ur Physik und Astronomie, Bochum, Germany}
 \author[0000-0002-7349-1109]{D. Guetta}
\affiliation{Physics Department, Ariel University, 40700 Ariel, Israel}
\author[0000-0003-2403-913X]{A. Lamastra}
\affiliation{INAF - Osservatorio Astronomico di Roma, Via Frascati 33, I-00078 Monte Porzio Catone (RM), Italy}
\author[0000-0002-2125-4670]{T. Shimizu}
\affiliation{Max-Planck f\"ur extraterrestrische Physik, Giessenbachstra{\ss}e 1, D-85748 Garching bei M\"unchen, Germany}


\date{\today}

\begin{abstract}

We present a comprehensive multi-messenger study of NGC 1068, the prototype Seyfert 
II galaxy recently associated with high-energy IceCube neutrinos. Various 
aspects of the source, including its nuclear activity, jet, outflow, and  
starburst region, are analyzed in detail using a multi-wavelength approach and relevant
luminosities are derived. We then explore its $\gamma$-ray and neutrino emissions 
and investigate potential mechanisms underlying these phenomena and their relations 
with the different astrophysical components to try to understand 
which one is responsible for the IceCube neutrinos. 
By first using simple order-of-magnitude arguments and then applying specific theoretical models, we infer that only the region close to the accretion disc around the supermassive black hole has both the right density of X-ray photons needed to provide the targets for protons to sustain neutrino production and of optical/infrared photons required to absorb the associated but unobserved $\gamma$ rays. We conclude by highlighting ongoing efforts to constrain a possible broad connection between neutrinos and active galactic nuclei, as well as future synergies between astronomical and neutrino facilities.\\
\end{abstract}

\section{Introduction}\label{sec:intro}

\begin{figure*}[t!]
    \centering\includegraphics[width=.55\textwidth]{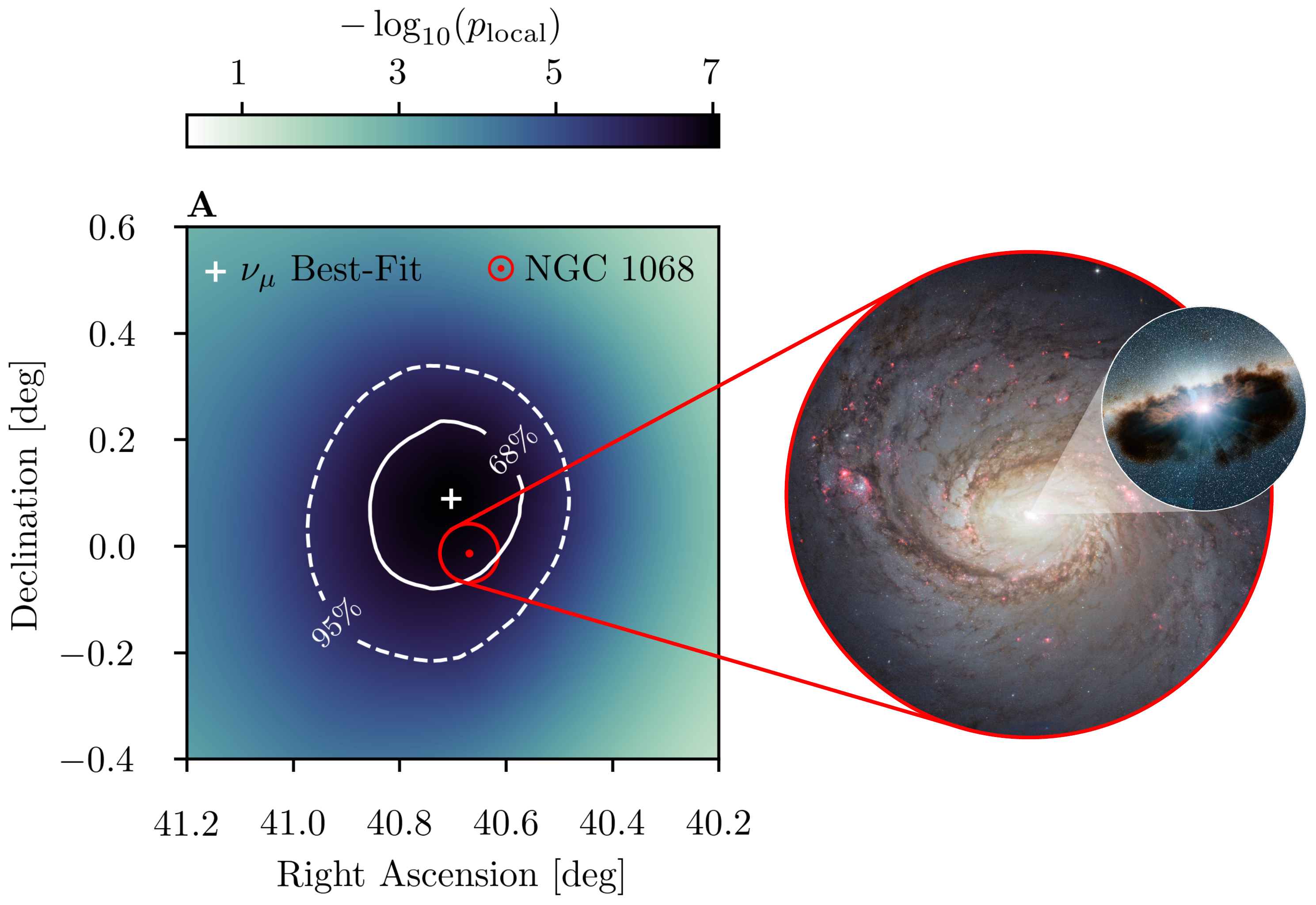}
    \caption{Significance map around the location of NGC~1068 from ref. \cite{Abbasi_2022}. The white cross indicates the most significant location in the neutrino sky and the white solid (dashed) contours show the 68\% (95\%) confidence regions around it. The red dot and the circle around it indicate the position and optical size of NGC~1068. On the right, the two insets show an image of the spiral galaxy and an illustration of its active nucleus surrounded by the dusty torus. Figure reproduced with 
permission from ref. \cite{Abbasi_2022} and NASA/JPL-Caltech.}
    \label{Fig:NGC_1068_hotspot}
\end{figure*}

Eleven years ago the IceCube Neutrino Observatory (\url{http://icecube.wisc.edu}) 
reported on the first high-energy (HE) astrophysical neutrinos of likely extragalactic origin
with energies in the TeV -- PeV range ($10^{12}$ - $10^{15}$ eV: \cite{Aartsen_2013,IceCube_2013}). 
This detection implies the existence of a class of astrophysical objects accelerating
 protons up to at least 10 -- 100 PeV, which then collide with other
 protons ($p - p$ collisions) or photons ($p - \gamma$ collisions). 
So far, however, searches performed by the IceCube collaboration have associated only the Galactic plane \cite{IceCube_2023} and two extragalactic objects with neutrinos at a significance larger than $\sim 3\,\sigma$. These are the blazar TXS\,0506+056 
at $z=0.3365$ (at the $3 - 3.5\,\sigma$ level: \cite{IceCube_2018a,IceCube_2018_b}) and
the prototype local Seyfert II galaxy NGC 1068 \cite{Abbasi_2022}. The latter paper has recently reported 4.2\,$\sigma$ evidence for an excess of TeV neutrinos from the direction of this source. 
As shown in Fig. \ref{Fig:NGC_1068_hotspot}, however, the 68\% confidence region around the most significant neutrino spot embeds the whole galaxy. Hence the resolution of the IceCube detector does not allow us to identify the region responsible for the cosmic-ray acceleration and the neutrino production processes, leaving open the question of the origin of these neutrinos. 

The main motivation of this review is to discuss the relevant astrophysical components
of NGC 1068, understand their nature, and constrain the origin of its neutrino 
emission from an observational, phenomenological, and theoretical multi-messenger perspective.

NGC 1068 is a very nearby spiral galaxy shown to have  ``bright emission
and absorption lines'' more than a century ago \cite{Fath_1909}. In 1943 
Carl Seyfert noticed that ``six extragalactic nebulae'' had ``high-excitation
nuclear emission lines [...] broadened, presumably by Doppler motion'' \cite{Seyfert_1943}.
The list included NGC 1068 and NGC 4151. Seyfert galaxies, as they came
to be known, are characterized by strong hydrogen, oxygen, neon, and other elements'
emission lines and get divided into two main classes, Seyfert Is and IIs. 
The latter are sources having only narrow lines, while the former have broad lines as well, where the 
boundary is drawn at 1,000 km s$^{-1}$ 
and is applied at the so-called full width at half maximum of the emission line. This
division is not simply semantic but, at the same time, it  turned out to be only apparent: 
Antonucci \& Miller in a seminal paper showed that the two classes represented the 
same types of sources seen at different angles \cite{Antonucci_1985}. NGC 1068, 
the prototype Seyfert II, turned out in fact to have broad lines but only in 
polarized light (due to scattering probably by free electrons). 
The picture which was put forward was one of dust in a doughnut-like configuration, 
the so-called ``torus'', surrounding the accretion disc on scales larger than those emitting 
the broad lines. In Seyfert IIs, which are observed edge-on with respect to the torus, only 
narrow emission lines, emitted further away than the broad lines and therefore less Doppler-broadened,  
could be observed in the optical spectrum. The inner regions of Seyfert IIs are 
therefore much more obscured and absorbed than Seyfert Is. This also led to the birth 
of ``unified schemes'', which unify apparently different but intrinsically similar classes
of active galactic nuclei (AGN). With the discovery of quasars \cite{Schmidt_1963}, in fact, 
Seyfert galaxies were shown to belong to the small fraction ($\approx 1\%$) of galaxies
in which matter falling onto the central black hole (BH) converts its gravitational energy
into radiation with the result that their nuclei become ``active'' and can easily outshine their
host galaxies \cite{Padovani_2017}. Much less than 10\% of all AGN become ``jetted'' 
\cite{Padovani_2011,Padovani_2017NatAs} and develop strong, long, relativistic jets, 
that is, structures in which matter is expelled at very high speeds, close to the speed of 
light, in relatively narrow beams. 

\begin{figure*}
    \centering\includegraphics[width=.8\textwidth]{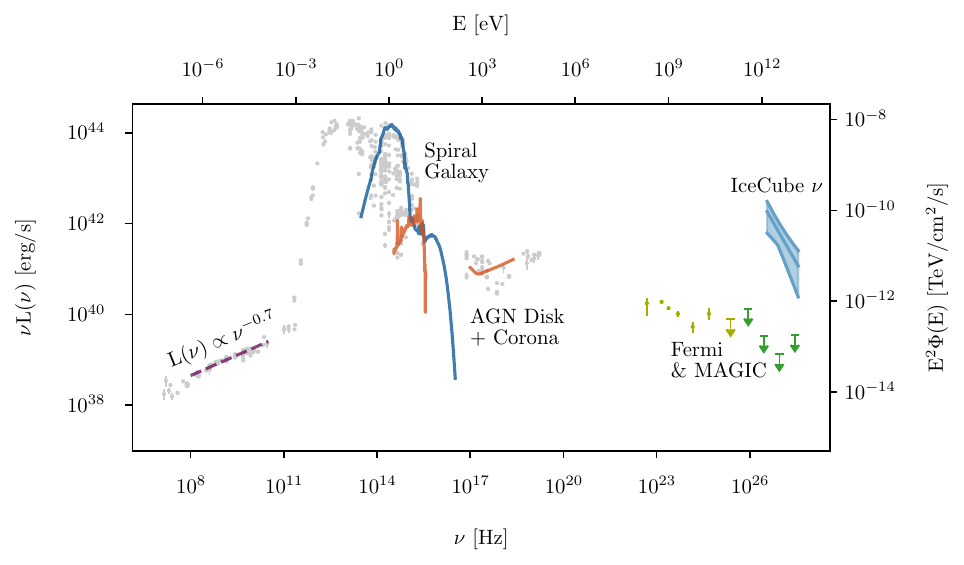}
    \caption{The integrated multi-messenger SED of NGC 1068, assembled using the
    SED builder at the SSDC (\url{https://tools.ssdc.asi.it/SED/}), showing the power emitted at different frequencies/energies; various components are highlighted. Note that the SED tool provides multiple values at the same frequency, which reflect the different resolutions used by the facilities which have studied the source on various spatial scales. For clarity we have kept mainly the highest flux values, which are associated with the largest scales, since the SED is dominated by the host galaxy given its relatively small distance.} 
    \label{Fig:NGC_1068_SED}
\end{figure*}

\section{The multi-messenger view}\label{sec:multi_view}

In this paper we assume a luminosity distance $D_{\rm{L}}=10.1$~Mpc (Methods \ref{app:facts}). 
This means that 1$^{\arcsec}$ corresponds to 48.9 pc and the  
BH mass \cite{Greenhill_1996} (which, being in this case $\propto r v^2$, scales 
like the linear size) can be estimated to be ${\rm M}_{\rm BH} 
= 6.7 \times 10^6~M_{\odot}$ (see discussion in ref. \cite{Wang_2020} on the robustness of this value), which implies an Eddington luminosity 
$L_{\rm Edd} \sim 8.4 \times 10^{44}$ erg s$^{-1}$ . 

Being so close, NGC 1068 can be spatially resolved into a number of components, 
all possibly relevant for neutrino production: 1. a starburst (SB) 
region in the spiral arms of its host galaxy; 2. a $\sim$ kpc jet; 3. a sub-kpc molecular
outflow; 4. the BH vicinity (the accretion disc and perhaps the so-called ``corona''). These can be studied
by using observations in different electromagnetic bands, which provide complementary
windows on the relevant physics. 

Before discussing the multi-wavelength properties, we first show the integrated 
multi-messenger spectral energy distribution (SED) in Fig. \ref{Fig:NGC_1068_SED}. 
A number of features have been highlighted: 1. the $\nu^{-0.7}$ radio spectrum; 
2. a template for the spiral host galaxy emission; 3. a template for the accretion 
disc plus X-ray corona emission; 4. the $\gamma$-ray and neutrino bands. 
Given the vicinity of this source its SED is largely due to the host galaxy, apart from
the X-ray band.

\begin{figure*}[ht]
\centering
\includegraphics[angle=-90,width=0.7\textwidth]{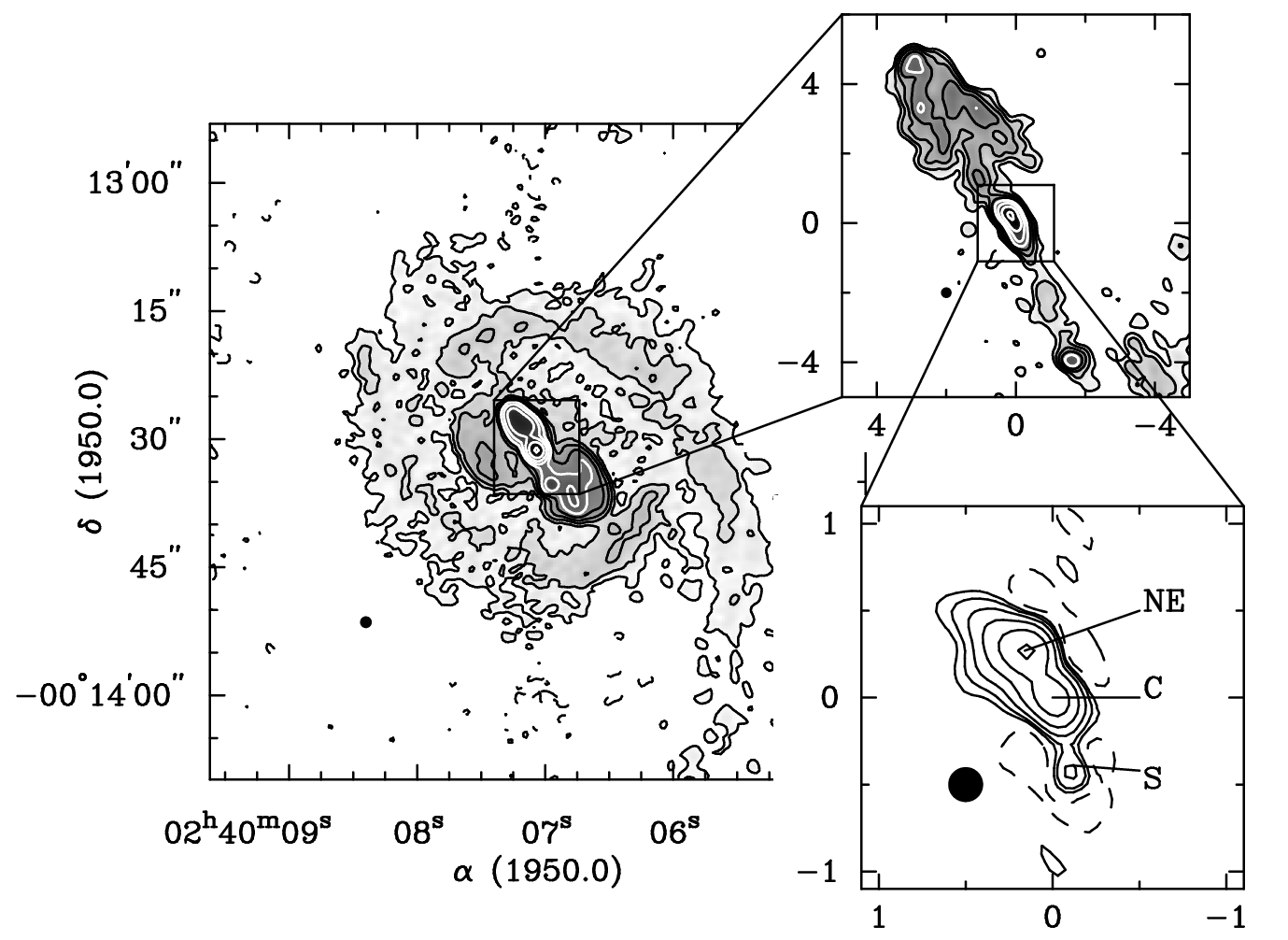}
\caption{Aperture synthesis image of NGC 1068 at 1.7 GHz (18 cm). 
Left: the SB disc; top right: the radio jet; bottom right: the 
central core. 1$^{\arcsec}$ corresponds to 48.9 pc. Figure reproduced with 
permission from ref. \cite{Gallimore_1996}, 
where one can find more details.} 
\label{Fig:NGC_1068_radio}
\end{figure*}

\subsection{The radio band: jet power}\label{sec:radio}

Fig. \ref{Fig:NGC_1068_radio} provides a beautiful overview of the inner few kpc
of NGC 1068 at 1.7 GHz \cite{Gallimore_1996}. The radio continuum image 
shows a $\sim 2$ kpc diameter SB disc (seen also through 
molecular tracers: \cite{Garcia_2014}), and what looks like a $\sim 500$ 
pc scale radio jet that terminates in lobes. 
The central 1$^{\arcsec}$ further 
resolves into a $\sim 50$ pc scale jet and compact radio knots (with the central engine believed to be close to component S). Note that the 
jet is almost two orders of magnitude shorter and much slower, for example, than the ``iconic'' 
M87 jet  ($< 0.05$c vs. $> 0.99$c \cite{Roy_2000,Pak_2019}), which 
implies that NGC 1068 is a ``non-jetted'' AGN following ref. \cite{Padovani_2017NatAs}. 
We also point out that ref. \cite{Fischer_2023} has suggested that the
velocities and trajectories of the radio flux peaks deviate from the expected behavior of emission originating from a relativistic radio jet. Instead, they seem to be linked to larger structures associated with the host gas reservoir. 
In other words, the extended radio structures 
seen in this AGN could also easily be by-products of local shocks. 

The total radio power at 1.4 GHz is $5.9 \times 10^{22}$ W Hz$^{-1}$ (using the 
NVSS radio flux density \cite{Condon_1998}), which makes NGC 1068 a moderately 
luminous Seyfert. Power per unit frequency is commonly used in radio astronomy: it can be converted to erg s$^{-1}$ at 1.4 GHz by multiplying by $1.4 \times 10^{16}$. The shape of the radio continuum (Fig. \ref{Fig:NGC_1068_SED})
indicates synchrotron emission, which does not help in figuring out its origin, 
as both SB and jets can produce it. However, one can derive the relative
contributions of the two components and then estimate from 
$L_{\rm jet,1.4 GHz}$ the {\it total} jet power,  $L_{\rm jet} = 10^{42.9\pm1.0}$ erg s$^{-1}$ (Methods \ref{app:radio_supp}). 
For comparison, TXS 0506+056, the blazar which 
is the only other extragalactic source associated with IceCube 
neutrinos, has $L_{\rm jet} 
\sim 10^{45} - 4 \times 10^{46}$ erg s$^{-1}$ based on SED modelling \cite{Ansoldi_2018}.

In principle both the jet and the SB region can be neutrino emitters through the
interaction of HE protons with photons and/or protons, as discussed in 
detail in Sections \ref{sec:neutrino}. 
 
\subsection{The sub-mm band: molecular outflow power}\label{sec:sub-mm}

AGN not only accrete matter to sustain their tremendous activity but many (possibly most) of them show also large-scale outflows/winds of matter driven by the central BH (for example, \cite{Harrison_2017,Cicone_2018} and references therein), which are both less collimated and slower than relativistic jets. These play a major role in galaxy evolution through the so-called AGN feedback (for example, ref. \cite{Fabian_2012} 
for a review). 
Such large-scale outflows may in fact quench star formation (SF) and starve the AGN through the lack of fuel. 
AGN outflows come in various ``phases'' characterized by different temperatures and compositions such as ionized, atomic, and molecular. These multiphase wide-angle winds have velocities ranging from $\sim 10^2$ km s$^{-1}$, typical of warm absorbers, up to semi-relativistic speeds $\sim 10^5$ km s$^{-1}$ for ultra fast outflows (UFOs), and are observed at different spatial scales (sub-pc to kpc) and ionization states \cite{Cicone_2018}. 

Most of the molecular gas in the interstellar medium (ISM) exists in the form of molecular hydrogen (H$_2$). 
H$_{2}$, however, is invisible
in the cold ISM where the majority of the molecular gas is located 
and where sub-mm observations are sensitive to (since its transitions can only be excited in warm regions probed by infrared [IR] or UV observations). 
Being the second most abundant molecule in the ISM,  
CO turns out to be the best tracer for the global molecular reservoir at sub-mm wavelengths. The Atacama Large Millimeter Array (ALMA), with its unprecedented sensitivity and spatial resolving power, is currently the most powerful tool 
to probe the molecular content in this band. 

A massive molecular outflow in NGC 1068 has been revealed by ALMA observations \cite{Garcia_2014,Impellizzeri+2019} with tracers including CO and hydrogen cyanide (HCN), and has been associated with AGN feedback. 
This outflow induces also molecular shocks at the circumnuclear disc (CND, $\sim$ 
300 pc scale: see for example refs. \cite{Viti+2014,Kelly+2017,Huang+2022}). 

The kinetic power related to a 
multi-conical outflow can be estimated from the following expression (refs. 
\cite{Maiolino_2012,Garcia_2014})

\begin{equation}
    L_{\rm kin} = {\frac{3~M_{\rm out}~v^3} {2~R_{\rm out}}} 
    \frac{\tan(\alpha)}{\cos(\alpha)^2}, 
    \label{eq:L_out}
\end{equation}

where $M_{\rm out}$ and $R_{\rm out}$ are the mass and radius of the outflow having velocity $v$ and $\alpha$ is the angle between the outflow axis and the line of sight. 
Using updated values for the distance and $\alpha \sim 
55^{\circ}$ (since the inclination is $i \sim 35^{\circ}$ from HyperLeda and $i$ and $\alpha$ are complementary if the outflow is co-planar with the galaxy disc)
and taking into account the uncertainties on the CO to H$_2$ conversion factor, $\alpha$, and $M_{\rm out}$, we derive what we consider a realistic estimate of $L_{\rm kin} = 10^{41.6\pm1.0}$ erg s$^{-1}$. 
The protons accelerated by the molecular
outflow shocks can generate $\gamma$-rays and neutrinos through the same process mentioned 
in Section \ref{sec:neutrino} (for example, ref. \cite{Lamastra_16}).
However, such energetic processes could also dissociate some of the molecules in the medium, and this effect is not taken into account in existing modelling scheme. 
 
\subsection{The near-IR band: the torus}\label{sec:NIR}

The Seyfert II nature of NGC 1068 (Section \ref{sec:intro}) implies the presence of a dusty torus, expected to be a few pc in size and thus surrounding the accretion disc but being within the CND. 
The torus imposes significant obscuration of 
the inner region of NGC 1068 along our line of sight. With an accretion disc temperature of $\approx 10^{4}-10^{6}$ K, the bulk of its emission is in the optical and UV bands and susceptible to absorption by dust. With high
enough column density through the torus ($N_H > 10^{24}\,\rm{cm}^{-2}$), even hard X-ray photons will get absorbed. All of this radiation is then reprocessed by nearby dust grains located close to the disc, and re-emitted in the IR range with a similar power (Section \ref{sec:X-ray}). 
The exact geometry and structure of the torus though is still a topic of debate and early observational work and recent modelling suggest a more complicated structure compared with the original homogeneous doughnut shape (for example refs. \cite{Ramos_2009,Markowitz_2014,Marinucci_2016}). With its expected small size, the torus can only be observed by interferometers with $\sim$ 100 m  baselines, and NGC 1068 has been a prime target for
the new IR interferometric instruments GRAVITY (K-band, 2.2 $\mu m$) and MATISSE (L, M and N-bands, 3 - 13 $\mu m$) at the ESO/VLTI in Cerro Paranal, Chile.

GRAVITY and MATISSE have now both confirmed the existence of, and resolved, the prominent dusty structure within the central few pc of NGC 1068.
GRAVITY first resolved the K-band emission into a thin, ring-like structure with a radius of 0.17 pc , which has been interpreted as the inner, hot edge of the torus with a thin disk morphology \cite{GRAVITY_2020}. MATISSE, in the nearby L, M, and N-bands has imaged the larger scale (up to 3 pc) colder dust into two distinct asymmetric structures with one extending in the polar directions \cite{2022Natur.602..403G}. Combining the two datasets, and the ability to fit SEDs over multiple bands, suggests that the K-band emission may arise mostly from the far side of the structure. The conclusion is that the torus is composed of a pc-scale inclined and optically thick inner disk and a prominent dusty outflow that connects to the outflow seen in molecular gas 
 (Section \ref{sec:sub-mm}). This lends support to the current theoretical picture of the torus as a disk plus wind complex and confirms the need for strong corrections to derive the true power of the central AGN in NGC 1068 to take absorption into account. 

\subsection{The X-ray band: the AGN power}\label{sec:X-ray} 

The X-ray band is key to study AGN because X-ray emission appears to 
be (near) universal and X-rays are able to penetrate through large column 
densities of gas and dust, particularly at higher energies ($\gtrsim 10$ keV: for example ref. \cite{Brandt2015}). 
Because of its rapid variability, X-rays can also resolve temporally the innermost 
parts of AGN, which cannot be resolved spatially with current technology (for example ref. \cite{DeMarco2022}).

The general consensus is that very hot ($T \approx 10^9$ K) 
electrons inverse Compton (IC) scatter the UV photons from the 
accretion disc thereby producing X-ray photons. By analogy with 
stellar sources, the X-ray radiating zone is also called the ``corona'' \cite{Liang_1979}. While its location has to be 
close to the disc, its detailed structure, plasma composition, and
origin, are not yet fully understood \cite{Alston2022}. In fact, many possible
corona structures have been proposed in the literature (Fig. 10
of ref. \cite{Lasota_2023}). Recent measurements obtained by the Imaging X-ray Polarization Explorer (IXPE) in Seyfert galaxies are challenging the commonly adopted lamp-post geometry, that is a point-like
corona illuminating a thin disc,  
favouring instead extended coronal geometries on/above the accretion disc plane \cite{Gianolli2023,Tagliacozzo2023,Ingram2023}.

The X-ray band is also important to estimate the bolometric luminosity, 
$L_{\rm bol}$, a key AGN parameter, whose determination requires a wealth of 
multi-wavelength data, not always available. This is particularly true
for Seyfert IIs where most of the AGN radiation is absorbed by dust and gas. While NGC~1068 is on average absorbed by a column density in excess of $10^{25}$ cm$^{-2}$, thus preventing a direct view of the intrinsic X-ray continuum, extensive monitoring campaigns with \textit{NuSTAR} detected an excess above 20 keV in some observations, suggesting unveiling events of the nucleus most likely due to a change in the absorption along the line-of-sight \cite{Marinucci_2016,Zaino2020}. This enabled the only direct measurement of the coronal X-ray emission of NGC~1068, $L_{\rm 2 - 10~keV}=3^{+3}_{-2}\times 10^{43}$ erg s$^{-1}$. 

One can then use the so-called X-ray bolometric correction $K_{\rm x}= L_{\rm bol}/L_{\rm x}$.  
Employing Table 1 of ref. \cite{Duras_2020} and $L_{\rm 2 - 10~keV}$ from the paragraph above we derive $L_{\rm bol,x} = 10^{44.5\pm0.4\pm0.3}$ erg 
s$^{-1}$ (where the first error term stems from the uncertainties on 
$L_{\rm 2 - 10~keV}$ while the second one is the spread of the correlation). 
Using a calibration based on the [OIV 25.89~$\mu$m] line luminosity,  
ref. \cite{Spinoglio_2022} has
estimated $L_{\rm bol,IR} = 10^{44.78}$ erg s$^{-1}$. This yields a 
logarithmic mean of $10^{44.7\pm0.5}$ erg s$^{-1}$, which is 
the value of $L_{\rm bol}$ we adopt in this paper, $\sim 400$ times higher than the all-flavour isotropic neutrino luminosity (Section \ref{sec:neutrino}). 
(This is fully consistent with
the logarithmic mean of the two extreme values given by ref. 
\cite{GRAVITY_2020}, that is $L_{\rm bol} = 10^{44.8\pm0.5}$ erg s$^{-1}$). We then derive $L_{\rm bol}/L_{\rm Edd} \sim 0.6$, considering  
the BH mass adopted in Section \ref{sec:multi_view}, which is
quite large for a local AGN.

\subsection{The $\gamma$-ray band: the SB power}\label{sec:gamma-ray}

A $\gamma$-ray source at the position of NGC~1068 was initially reported in 2010 in the first {\it Fermi}-LAT catalogue (1FGL) \cite{1FGL} and then associated  with NGC 1068  \cite{Lenanin_2010}. Since then, NGC 1068 has been included in every {\it Fermi}-LAT catalogue, up to the 4FGL-DR4, at an increasing level of significance \cite{4FGL}. Its average power in the 0.1 -- 100 GeV range over twelve years is $L_{\gamma} = 10^{40.92\pm0.03}$ erg s$^{-1}$, $\sim 15$ times smaller than the all-flavour isotropic neutrino luminosity (albeit in a different energy range: Section \ref{sec:neutrino})

In the very-high energy (VHE) band, the source was observed by ground-based Cherenkov telescopes \cite{Aharonian2005,Acciari_2019,Albert_2021} with no detection.
The most stringent constraints on the VHE flux were placed by a 125-hour long observation performed by the Major Atmospheric Gamma-ray Imaging Cherenkov (MAGIC) telescopes, which  provided a 95\,\% confidence level upper limit to the $\gamma$-ray flux above 200\,GeV of 5.1 $\times 10^{-13}$ cm$^{-2}$ s$^{-1}$ \cite{Acciari_2019}.

The GeV $\gamma$-ray emission of NGC 1068 has been interpreted as due to SF activity \cite{Ajello_2020}, which through the creation of stellar remnants, like pulsars, pulsar-wind nebulae, and supernova remnants, leads to the acceleration of cosmic rays (CRs). Their interaction produces $\gamma$-rays. In SB galaxies like NGC 1068 (Methods \ref{app:radio_supp}) most of 
this emission is of hadronic origin and produced via $p - p$ interaction and the decay of neutral pions \cite{Peretti_2019}.
SF galaxies are known to follow an $L_{\gamma}-L_{\rm IR}$ correlation where indeed both luminosities are tracers of SF activity. The agreement of NGC 1068 with this correlation and the lack of variability at $\gamma$ rays (for example refs. \cite{Ajello_2020, ajello2023}) support the interpretation that the majority of the emission is related to SF activity.

A contribution to the $\gamma$-ray emission may in principle come from a UFO in NGC 1068. Indeed, ref. \cite{Ajello_2021} has reported the detection of $\gamma$-rays  from a sample of nearby AGN with a UFO. Adopting the scaling relations in ref. \cite{Ajello_2021} and estimating $L_{\rm kin} \lesssim 10^{43}$\,erg s$^{-1}$ from the $L_{\rm kin} - L_{\rm bol}$ relationship of ref. \cite{Fiore_2017}, one can derive a range for the UFO $\gamma$-ray luminosity between $10^{39.6\pm1.0}$\,erg s$^{-1}$ and $10^{41.2\pm1.0}$\,erg s$^{-1}$, where the uncertainties take into account the errors in the scaling relations of ref. \cite{Ajello_2021}.
Note that, while the UFO may explain (and actually overshoot) the $\gamma$-ray luminosity of the source, there is no robust evidence that such UFO exists in NGC 1068 as UFOs are nearly impossible to detect in the X-rays in heavily obscured 
AGN, where direct observation of the intrinsic nuclear continuum in the Fe K band is hindered \cite{Tombesi2010}.

On the other hand, as shown in Section~\ref{sec:sub-mm}, NGC 1068 exhibits a massive molecular outflow with an
outflow rate $\approx 38~M_{\odot}$\,yr$^{-1}$ (Section \ref{sec:sub-mm}). $\gamma$-ray emission has  been observed from a sample of nearby galaxies (including NGC 1068) with a molecular outflow \cite{McDaniel_2023}. 
Contrary to the UFO case, however, no correlation was found between $L_{\rm kin}$ and $L_{\gamma}$, which makes it impossible
to estimate any contribution from the molecular outflow to the $\gamma$-ray emission. 

$\gamma$-ray emission from the jet is very likely negligible since NGC 1068's 
jet is not as strong and fast (Section \ref{sec:radio}) as the relativistic jets 
in the sources typically detected by {\it Fermi}-LAT, that is blazars, which are also very variable. 
In fact, assuming the same $L_{\gamma}/L_{{\rm 1.4\,GHz}}$ 
ratio as the blazar TXS 0506+056 \cite{Paiano_2023}, where both
powers are jet-related, $L_{\rm \gamma,jet}$ in NGC 1068 should be 
$\sim 70$ times higher than observed (or $\sim 7$ times
in the less extreme case of an M87-like jet). 

\subsection{The TeV neutrino band}\label{sec:neutrino}

The IceCube result was the outcome of a new and improved analysis of nine years ($2011-2020$) of neutrino event candidates that were updated to newer processing and detector calibration \cite{Abbasi_2021_PhRvD}. 
The search that led to the detection of significant neutrino emission from NGC~1068 tested 110 
$\gamma$-ray emitters \cite{Fermi_4fgl_dr2_2020} defined a priori and located 
in a declination range covering the Northern sky, $-3^{\circ}<\delta<81^{\circ}$. The selected objects included 95 blazars, 5 AGNs, 9 other types of galaxies, and 1 Galactic source \cite{Abbasi_2022}.
Dedicated searches in the direction of pre-defined source candidates are not new to the IceCube analyses (see, for example, ref. \cite{IceCubeApJ2009}). The selection method utilized to identify the 110 interesting candidates used by ref. \cite{Abbasi_2022} was the same as in a previous IceCube analysis \cite{IceCubePRL2020} that reported a 2.9\,$\sigma$ excess from the direction of NGC~1068. The most significant object in the list was again NGC~1068, and indeed the most significant location identified by the scan of the Northern hemisphere was found only 0.11$^{\circ}$ away from this source (Fig. \ref{Fig:NGC_1068_hotspot}). The muon-neutrino energy flux from NGC~1068 at 1~TeV under the assumption of a single power-law has a best-fit normalization $\Phi^{1\rm{TeV}}_{\upnu_{\upmu}+\bar{\upnu}_{\upmu}}=(5.0\pm2.1_{\rm{stat+sys}})\times10^{-11}\,\rm{TeV}^{-1} \rm{cm}^{-2} \rm{s}^{-1}$ and a best-fit spectral index $\Gamma=3.2\pm0.3_{\rm{stat+sys}}$, valid in the energy range between 1.5 and 15~TeV. This flux corresponds to an equivalent all-flavour isotropic neutrino luminosity $L_{\upnu}=10^{42.1\pm0.2}$ erg s$^{-1}$ in the same energy range.

\begin{table*}
\caption{Measured powers.}
 \begin{center}
 \begin{tabular}{cccccc}
   \hline
    $L_{\rm radio}$ & $L_{\rm FIR}$ & $L_{\rm x}$ & $L_{\gamma}$ & $L_{\nu}$ & $L_{\rm Edd}$\\
    (1.4 GHz) &  (8 -- 1000 $\mu$m) & (2 -- 10 keV) & (0.1 -- 100 GeV) & (1.5 -- 15 TeV) &\\
\hline
    $10^{38.9}$ & $10^{44.6\pm0.1}$ & $10^{43.4\pm0.3}$ & $10^{40.92\pm0.03}$ & $10^{42.1\pm0.2}$ &
    $10^{44.9\pm0.3}$~$^a$\\
  \hline
  \end{tabular}
  \end{center}
    \footnotesize {~~~~~~~~~~~~~~~~~~~~~~~~~~~~~~~~~All powers in erg s$^{-1}$. $^a$: factor of 2 uncertainty assumed on ${\rm M}_{\rm BH}$.}  
 \label{tab:measured_powers}
\end{table*}

\begin{table*}
\caption{Derived powers.}
 \begin{center}
 \begin{tabular}{cccc}
   \hline
    $L_{\rm kin}$ & $L_{\rm jet}$ & $L_{\rm bol}$ & $L_{\rm bol}/L_{\rm Edd}$ \\
\hline
    $10^{41.6\pm1.0}$ & $10^{42.9\pm1.0}$ & $10^{44.7\pm0.5}$ & $10^{-0.3\pm0.6}$ \\
  \hline
  \end{tabular}
  \end{center}
  \footnotesize {~~~~~~~~~~~~~~~~~~~~~~~~~~~~~~~~~All powers in erg s$^{-1}$ apart from the Eddington ratio.}  
 \label{tab:derived_powers}
\end{table*}

\begin{table*}
\caption{Estimated $\gamma$-ray and neutrino powers.}
 \begin{center}
 \begin{tabular}{cclc}
   \hline
    Component & Scale & ~~~~~~~$L_{\gamma}$ & $L_{\nu}$ \\
   &  & (0.1 -- 100 GeV) & (1.5 -- 15 TeV) \\
\hline
   Star formation & $>$ kpc & $\sim 10^{40.9}$ & $\lesssim 10^{40.1}$ \\
   Jet & $\sim$ kpc & $< 10^{41.7}$ (M87-like) & $<  10^{40.9}$ \\
   Outflow (UFO) & $\sim$ pc & $< 10^{41.2}$  & $<  10^{40.4}$ \\
   BH vicinity & $\sim$ 0.03 mpc ($\sim 50~R_s$) & ~~~~? & ? \\
  \hline
       &  ~~~~Total  & $\lesssim 10^{41.9}$  & $\ll 10^{41.1}$    \\
       &  Observed  & ~~~$10^{40.92\pm0.03}$  & ~~~~~~~~~$10^{42.1\pm0.2}$    \\
  \hline       
  \end{tabular}
  \end{center}
  \footnotesize {~~~~~~~~~~~~~~~~~~~~~~~~~~~~~~~~~All powers in erg s$^{-1}$; $R_s$ is the Schwarzschild radius.}  
 \label{tab:neutrino_powers}
\end{table*}

High energy neutrinos are generated in the decay of charged pions, which are in turn produced when CRs interact with ambient matter ($p - p$) or low-energy radiation (photohadronic [$p - \gamma$] interactions).
The final products of the decay chain of a charged pion
\begin{eqnarray}
\pi^{\pm} &\rightarrow& \mu^{\pm} + \nu_{\mu} (\bar{\nu}_{\mu}) \\
\mu^{\pm} &\rightarrow& e^{\pm} +\bar{\nu}_{\mu} (\nu_{\mu}) + \nu_e (\bar{\nu}_e)
\end{eqnarray}
are four stable particles: one electron (or positron) and three neutrinos and antineutrinos \cite{berezinskii1990,gaisser2016}.

Neutral pions are also produced in $p - p$ and $p - \gamma$ interactions, approximately in the same amount as charged ones, decaying into two $\gamma$ rays, $\pi^0 \rightarrow \gamma + \gamma$.
The energy of the pions (both charged and neutral) is shared roughly equally among the final products of the decay, and this has two important consequences:
1. a flux of neutrinos from astrophysical sources is unavoidably accompanied by a $\gamma$-ray flux $F_{\gamma} \sim 2 \times F_{\nu}$; 2. the energy of a neutrino ($E_{\nu}$) and of a photon ($E_{\gamma}$) generated in CR interactions are proportional to the CR proton energy $E_p$ and related by $E_{\gamma} \sim 2 \times E_{\nu}$ \cite{berezinskii1990,gaisser2016}. 

\section{Relevant powers and the case for a ``hidden'' source scenario}\label{sec:powers} 

Tables \ref{tab:measured_powers} and \ref{tab:derived_powers} summarise
the relevant powers discussed in this paper, where we distinguish between 
``measured'' (radio, far-infrared [FIR], X-ray, $\gamma$-ray, neutrino, and Eddington)
and ``derived'' (outflow, jet, bolometric, and Eddington ratio) 
values respectively. 

Having examined all available data, both in the 
electromagnetic and neutrino bands, we now address
the question of the origin of the neutrino emission. 
Our starting point is the fact that NGC~1068 is much weaker 
in TeV photons than in neutrinos (see upper limits in 
Fig.~\ref{Fig:NGC_1068_SED} and Sections \ref{sec:gamma-ray} and \ref{sec:neutrino}). 

Table \ref{tab:neutrino_powers} puts this statement into perspective 
by estimating the {\it maximum} neutrino power
that can be associated to the various components. 
This was done by making three extreme assumptions: 1. $\gamma$-ray 
photons are fully hadronic (hence  $L_{\nu} \sim L_{\gamma}/2$); 2. $L_{\gamma}$ is as 
large as possible; 3. the conversion between the {\it Fermi} and the IceCube bands 
is done assuming $\Gamma_\nu = 2$ (while the observed one is $\Gamma_\nu = 3.2$),
which implies, together with assumption n. 1, that $L_{\nu} (1.5 - 15~ {\rm TeV}) 
\sim L_{\gamma} (0.1 - 100~{\rm GeV})/6$. Any deviation from these premises will
result in lower neutrino powers. While the SF-related
$L_{\gamma}$ is the observed one, 
we have maximized the $\gamma$-ray emission associated with the two other components by 
assuming an absorbed M87-like jet power, 
which is totally unphysical, and the maximum possible value 
for a UFO for the outflow (Section \ref{sec:gamma-ray}). The resulting
neutrino power has then to be $\ll 10^{41.1}$ erg s$^{-1}$, that is more than a factor of ten less (realistically at least two orders of
magnitude using only the SF component) than the IceCube value. 
Since in a photohadronic scenario neutrinos
are associated with a twice as high and twice as energetic $\gamma$-ray flux, Figure \ref{Fig:NGC_1068_SED} implies that the photons related to neutrino production have an intrinsic flux
$\sim 40$ times higher than the MAGIC upper limits (Section \ref{sec:gamma-ray}) and therefore need to be  
totally absorbed. 

We now evaluate the characteristics that the emission region must possess to meet the criteria derived above.
We focus here on the $p - \gamma$ scenario as it provides a very natural explanation for both the neutrino flux and the low level of gamma-ray emission. We remind the reader, however, that if the gas density is large enough, also a $p-p$ origin of high-energy neutrinos cannot be ruled out.

The $p - \gamma$ cross section is characterized by a pronounced peak just above the threshold for pion production \cite{mucke1999}.
For well behaved photon spectra, 
most $p - \gamma$ interactions 
occur near threshold and result in the production of a single pion, either charged or neutral (multiple-pion production takes over at larger energies). 
Moreover, the incident proton loses a fraction $m_{\pi}/m_p$ (the ratio of the pion to proton mass) of its energy in each interaction  and the characteristic energy of each neutrino is $E_{\nu} \sim (m_{\pi}/4 m_p) E_p \sim 0.04 \times E_p$, where the factor of four represents the number of stable particles resulting from the interaction \cite{berezinskii1990}. 

The neutrinos observed by IceCube are therefore generated by CR protons of energy $\sim 40 - 400$~TeV. 
Due to the threshold condition, in order to actually produce neutrinos, protons must interact with photons of energy  

\begin{equation}
\epsilon_{\gamma}^{th} \gtrsim 2 ~(E_p/40~{\rm TeV})^{-1}~{\rm keV},
\label{eq:eps_th}
\end{equation}

 that is in the X-ray domain. 

Remarkably, neutrino production in a photon rich environment would also explain the flux suppression in $\gamma$ rays, due to the onset of pair production in the ambient radiation field, provided that the threshold condition $E_{\gamma} \epsilon_{\gamma} > (m_e c^2)^2$ is satisfied \cite{gould1967,aharonian2004}. 
This translates into a minimum energy 

\begin{equation}
\epsilon_{\gamma} > 0.26 ~(E_{\gamma}/{\rm TeV})^{-1}~{\rm eV},
\label{eq:eps_gam}
\end{equation}

that is in the IR domain ($\lambda < 4.8~\mu m$) and above (energy wise). 

The conditions required by equations (\ref{eq:eps_th}) and (\ref{eq:eps_gam}) are easily satisfied very
close to the supermassive BH at the centre of the AGN, a region which provides X-ray and optical/IR photons in
abundance \cite{inoue2019}. In particular, the X-ray corona (Section \ref{sec:X-ray}) has been singled out as the region
responsible for neutrino emission (Section \ref{sec:theor_models}).

Electron-positron pairs generated this way will IC scatter ambient photons up to the $\gamma$-ray domain, thereby initiating an electromagnetic cascade, sustained by alternate IC and pair production processes.
The cascade reaches its natural end when the typical energy of photons equals the threshold for pair production: $E_{\gamma}^{th} = (m_e c^2)^2/\epsilon_{\gamma}$.
The electron/positron generated in the last pair production event will have an energy $E_e^{min} \sim E_{\gamma}^{th}/2$, and will IC upscatter ambient photons to an energy $E_{\gamma}^{esc} = (4/3) (E_e^{min}/m_e c^2)^2 \epsilon_{\gamma} = E_{\gamma}^{th}/3$.
Photons of such energy are unable to pair-produce in the ambient photon gas, and will freely escape the region \cite{berezinskii1990}.

The spectrum of ambient photons in the AGN corona extends well into the keV range. These X-rays, which are the targets for neutrino production, are also the reason why the AGN corona remains opaque for $\sim$ GeV $\gamma$-rays, given that the pair production cross section is much larger than that of pion production. Runaway photons escaping from this environment need to have at most MeV energies,
$E_{\gamma}^{esc} \sim 45~ (\epsilon_{\gamma}/2~{\rm keV})^{-1}$~MeV, which are, to date, in a very poorly explored {energy range}.
Future MeV missions, such as, for example, AMEGO-X \cite{Caputo:2022xpx} and e-ASTROGAM \cite{deangelis2018}, 
are therefore of paramount importance, as they might be the only instruments capable of revealing the electromagnetic counterpart of neutrino sources similar to NGC~1068. 

The same conclusion is reached also in a scenario where the electromagnetic cascade is completely suppressed due to the presence of a magnetic field \cite{Gabici_2007}. 
If the ambient magnetic field is strong enough, multi TeV CRs would produce TeV electrons and such electrons would cool through synchrotron emission rather than IC/pair production processes.
For a kiloGauss field, that might be found in AGN coronae
(Section \ref{sec:AGN_corona}), the energy of the synchrotron photons falls in the MeV domain, $\epsilon_{syn} \sim 20 (B/{\rm kG}) (E_e/{\rm TeV})^2$~MeV (see, for example, ref. \cite{Murase2022}).

\section{Nailing down the hidden source case}
\label{sec:theor_models}

The existence of neutrino sources not associated to any (except for MeV) photon counterpart (and therefore named hidden sources) was first predicted in the seventies in a number of pioneer papers \cite{berezinsky1977,eichler1979,silberberg1979}.
NGC~1068 is the first such source to be detected, and this revived the interest in modelling neutrino production in environments opaque to high energy radiation (for example \cite{Inoue2020,Murase2020}).

We now expand on Section \ref{sec:powers} and establish theoretical connections between the primary candidate astrophysical sites, particle acceleration, and interactions, summarizing the multi-messenger implications for NGC~1068.

\subsection{Star-forming region}
SF regions and SBs are powerful cosmic-ray factories due to a high 
supernova rate which in turn is a result of an enhanced SFR.
Being the seed of star formation, a high gas density ($n \gtrsim 10^2 \, \rm cm^{-3}$) is typically found in SF environments.  
Such a density is high enough to make the inelastic $p - p$ interaction rate comparable with or more efficient than the escape rate. 
Therefore, copious production of $\gamma$-rays and HE neutrinos is expected (Section \ref{sec:neutrino}). 
The SF environment possesses also a strong FIR dust emission ({with radiation energy density} $U_{\rm RAD}\gtrsim 10^3 \, \rm eV \, cm^{-3}$) resulting from the efficient absorption of starlight which is reprocessed at lower frequencies. 
This photon background is responsible for efficient $\gamma \gamma$ absorption of $\gtrsim 10$ TeV photons while those at lower energies are unabsorbed 
(equation \ref{eq:eps_gam}).

The GeV $\gamma$-ray luminosity of NGC~1068 is roughly compatible with what can be expected from the FIR-$\gamma$-ray correlation 
(Section \ref{sec:gamma-ray}). 
Therefore, the SB is an interesting candidate for explaining the bulk of the observed GeV $\gamma$-rays.
NGC~1068 is characterized by a SB ring of radius $\approx 1 \, \rm kpc$ and a CND. 
Ref.~\cite{Yoast-Hull_2014} found that the inner CND cannot explain the $\gamma$-ray data without overproducing the radio constraints. 
However, ref.~\cite{Eichmann_2022} pointed out that the SB ring features the right conditions to be consistent with the multi-wavelength observations, but cannot be the main source of the IceCube neutrino flux. 
Finally, ref.~\cite{Eichmann_2022} and ref.~\cite{ajello2023} reported that NGC 1068's 20\,MeV - 1~TeV spectrum 
can be interpreted as the sum of two hadronic components: a highly obscured one related to the AGN and responsible for the $\lesssim$ 500\,MeV emission and another one produced in the SB ring by SF activity.

\subsection{AGN outflow}\label{sec:theor_outflow}
AGN outflows/winds (Sections \ref{sec:sub-mm} and \ref{sec:gamma-ray}) may be launched from the accretion disc by  several mechanisms involving thermal, radiative, and magnetic processes. 
During propagation, AGN-driven winds, similarly to other diverging flows, are expected to develop a structure characterized by an inner wind termination shock and an outer forward shock.
As discussed in the following paragraphs, both have been proposed as possible sites for particle acceleration through diffusive shock acceleration. %

Particle acceleration at the forward shock of the AGN-driven molecular wind observed in NGC~1068 is strongly constrained by the upper limits on the $\gamma$-ray flux in the VHE band obtained by the MAGIC telescopes (see Section
\ref{sec:gamma-ray}). 
This outflow expands at a large distance in the host galaxy and $p - p$  inelastic collisions are the main production mechanism for $\gamma$-rays and neutrinos. 
This model predicts a maximum neutrino event rate of $\sim$ 0.07 yr$^{-1}$, which is much lower than that observed by IceCube \cite{Lamastra_16}. 

{Parsec and sub-pc scale fast winds, such as UFOs, could also be present in the nuclear region of NGC~1068. However, the Compton-thick nature of such a nuclear region makes a detection of such winds extremely challenging} (Sections \ref{sec:X-ray} and \ref{sec:gamma-ray}).
{In a small-scale wind such as a UFO} $p - p$ and {$p - \gamma$ interactions  
could account for the $\gamma$-ray emission observed in the {\it Fermi}-LAT band, while their contribution to the IceCube neutrino flux would be limited to $<$10\% \cite{Peretti_23}. 

Finally, a scenario in which neutrinos are produced in the inner regions of a failed wind (that is, a wind that never reaches the local escape velocity) was also investigated \cite{Inoue_22}. 
In this context neutrinos are mainly generated via $p - \gamma$ interactions with the AGN radiation field, and the predicted neutrino flux %
is as high as the one observed by IceCube above 1 TeV. 
This scenario shares some similarities with the AGN corona models 
(Section \ref{sec:AGN_corona}), 
such as the spatial scale of the neutrino production site and the vicinity to the emission region of the target X-ray AGN photon field.
This system is optically thick for GeV -- TeV $\gamma$-rays. As for the corona case,  
in order to reproduce the observed $\gamma$-ray spectrum in the {\it Fermi}-LAT band a separate external emission region must be postulated, that is a larger scale wind or the SB  
{ring}. 

\subsection{Jet}

NGC~1068 has a weak and slow jet with a kinetic luminosity $L_{\rm jet} \sim 10^{43} \, \rm erg \, s^{-1}$ (see Section~\ref{sec:radio}).
Ref.~\cite{Lenanin_2010} proposed a leptonic scenario 
{for the observed GeV flux} 
%
with an artificial broken power-law spectrum of relativistic electrons that extends up to $0.5\,\text{TeV}$ and IC scatters IR photons from the torus within pc-scale blob structures. However, the lack of time-variability of the $\gamma$-ray flux (Section \ref{sec:gamma-ray}) and the need for a distinct acceleration-cooling scenario, as pointed out by ref. \cite{Salvatore_2024}, make a leptonic jet interpretation unlikely.
{In this context, a hadronic counterpart} is disfavored for explaining a sizable fraction of the IceCube neutrino flux unless the proton-dominated emission is spatially and energetically decoupled from the associated leptonic {component} and localized in an optically thick region. Such a scenario is discussed by ref.~\cite{Fang_2023} for certain knot structures within the jet, where a peak in the $\gamma$-ray energy spectrum between $100\,\text{GeV}$ and $1\,\text{TeV}$ as well as a dominant neutrino flux contribution above $\sim 10$ TeV is predicted if the proton energy spectrum is sufficiently hard. The possible CR production site in the head of the jet \cite{Michiyama_2022} yields another potential hadronic contribution to the $\gamma$-ray energy spectrum that can become dominant above a few GeVs \cite{Salvatore_2024}. But in any case the entire spectrum of the observed $\gamma$-rays cannot be explained by any of these structures within the jet.

\subsection{AGN corona} 
\label{sec:AGN_corona}

As discussed in Section~\ref{sec:powers},
the large discrepancy between neutrino and $\gamma$-ray powers (see also Fig. \ref{Fig:NGC_1068_SED} and Tables \ref{tab:measured_powers} and \ref{tab:neutrino_powers}) 
requires a high opacity for $\gamma$-rays. 
This is accommodated rather naturally if the acceleration/interaction region is identified as the corona of the supermassive BH (Sections \ref{sec:X-ray} and \ref{sec:powers}; see also ref. \cite{Murase2020}). The high temperature of the gas expected in the corona leads us to think that shock waves in this region should be rather weak, thereby leading to steep spectra of accelerated particles.

In this context, refs.~\cite{Murase2020} and \cite{Inoue2020} proposed respectively stochastic acceleration in turbulence (second order Fermi process) and diffusive shock acceleration (first order Fermi process) as main acceleration mechanisms. 
Both scenarios are able to produce a strong neutrino signal while the $\gamma$-ray counterpart is efficiently absorbed and reprocessed in the MeV band (see also refs. \cite{Murase2022,Eichmann_2022,Inoue_22}).

AGN coronae are believed to extend for less than $\sim 10^2$ $R_s$ 
and are characterized by an intense X-ray radiation field associated to the accretion activity. 
Some simple estimates can illustrate the rough properties of the corona: its radius can be parametrized as $R_c=\eta R_s$ (where $\eta\sim 10-100$), so that, assuming approximate virial balance between gravitational, thermal and magnetic energy, the proton temperature of the corona becomes $k_B T_p = m_p c^2/6\eta$, that is 
$T_p \sim 6.1 \times 10^{10} (\eta/30)^{-1}$ K. Note that this is about an order of magnitude larger than that of thermal electrons as estimated from the cut-off in the observed X-ray spectra. Although some level of thermalisation between protons and electrons is expected, the time scale for full equilibration between the two species is exceedingly long \cite{Ghisellini_1993,Fabian_2015,Fabian_2017}, and it is reasonable to expect that protons in the corona are much hotter than electrons.

The magnetic field in the corona can also be estimated as:
\begin{equation}
    B=\sqrt{\frac{4\pi\rho_p}{3\eta \beta_P}}c = 2.2 \times 10^3 \rm G ~(\eta/30)^{-1} M_7^{-1/2}\tau_T^{1/2}\beta_P^{-1/2},
\end{equation}
where $\rho_p=n_p m_p$ is the proton mass density and $\beta_P=8\pi n_p k_B T_p/B^2$ is the ratio between the magnetic and thermal energy densities. Here we introduced the Thomson opacity of the corona $\tau_T=n_p\sigma_T R_c$, $M_7$ as the BH mass in units of $10^7M_\odot$, and we have used the results previously obtained for the gas temperature. From the definition
of the Alfv\'en speed ($v_A$) one can see that in the corona
 $v_A/c_s\sim \beta_P^{-1/2}$ and $v_A=c/\sqrt{3\eta\beta_P}$,
 where $c_s$ is the sound speed. The fact that $v_A$ is close to the speed of light is an important ingredient to make turbulent particle acceleration efficient.

Interestingly, the spectrum of high energy neutrinos measured from the direction of NGC 1068 is very steep, $\propto E_\nu^{-3.2}$. This piece of information is crucial to estimate the total luminosity in the form of accelerated particles necessary to power the neutrino signal. As discussed above, neutrino production at $E_\nu \sim 1$ TeV in $p - p$ and $p - \gamma$ reactions requires a proton with $E\sim 25$ TeV. Assuming that protons dump all their energy near the black hole, the observed neutrino luminosity requires a proton luminosity at energies $>25$ TeV $\sim 10^{42}$ erg s$^{-1}$. Extrapolating this down to $1$ GeV assuming the spectrum of the IceCube neutrinos, translates into $\sim 5 \times 10^{46}$ erg s$^{-1}$.  
This number underestimates the real luminosity by at least a factor $\sim 2$ (because of particles with $E<1$ GeV). It is clear that if the steep spectrum of parent protons were to extend to low energies, then both $L_{\rm Edd}$ (Table \ref{tab:measured_powers}) and $L_{\rm bol}$ (Table \ref{tab:derived_powers}) would be exceeded (note also that in these estimates we are neglecting the energy required to feed the thermal and magnetic energy of the corona). 

The most natural explanation for the steep neutrino spectrum is that it reflects the cutoff region of the parent protons, while the lower energy spectrum should be hard enough to avoid or alleviate problems with energetics. 

These conclusions reflect on considerations concerning the specific acceleration process that is responsible for particle energisation. Shocks in the environment associated with the corona are likely to have low Mach number, given the fact that the free fall velocity and the sound speed are very close to each other, hence the spectrum of accelerated particles is expected to be much steeper than $E^{-2}$. {\bf (Alternatively, the entire corona could be the downstream region of a strong shock, so that particles could be accelerated elsewhere and eventually advected into the corona.)}
As discussed above, while this result can be in line with the observed steep neutrino spectrum, it also creates severe problems with global energetics. On the other hand, the high value of the Alfv\'en speed in the corona suggests that stochastic acceleration may be at work. Such a process may naturally produce hard spectra with a cutoff energy consistent with the observed neutrino spectrum. This possibility is discussed in more detail in Methods \ref{app:accel}.

It is useful to point out that while magnetohydrodynamic (MHD) simulations of the region near the BH do not show evidence of shocks \cite{Porth2023}, they do suggest that a hot corona is formed above the accretion disc \cite{Jiang2019}, lending support to the picture outlined above.

Recently ref. \cite{Fiorillo_2023} (see also ref.~\cite{Mbarek_2024}) explored relativistic magnetic reconnection as an alternative acceleration mechanism responsible for the injection in the corona of a hard proton spectrum ($\propto E^{-1}$) featuring a break at about $\sim$ 25 TeV. Such a scenario can be a viable alternative and/or a complement to the stochastic acceleration in providing the non-thermal proton population the necessary energetic to power the observed neutrino flux.

\section{Outlook}\label{sec:outlook}

\begin{figure*}[t]
    \centering\includegraphics[width=.7\textwidth]{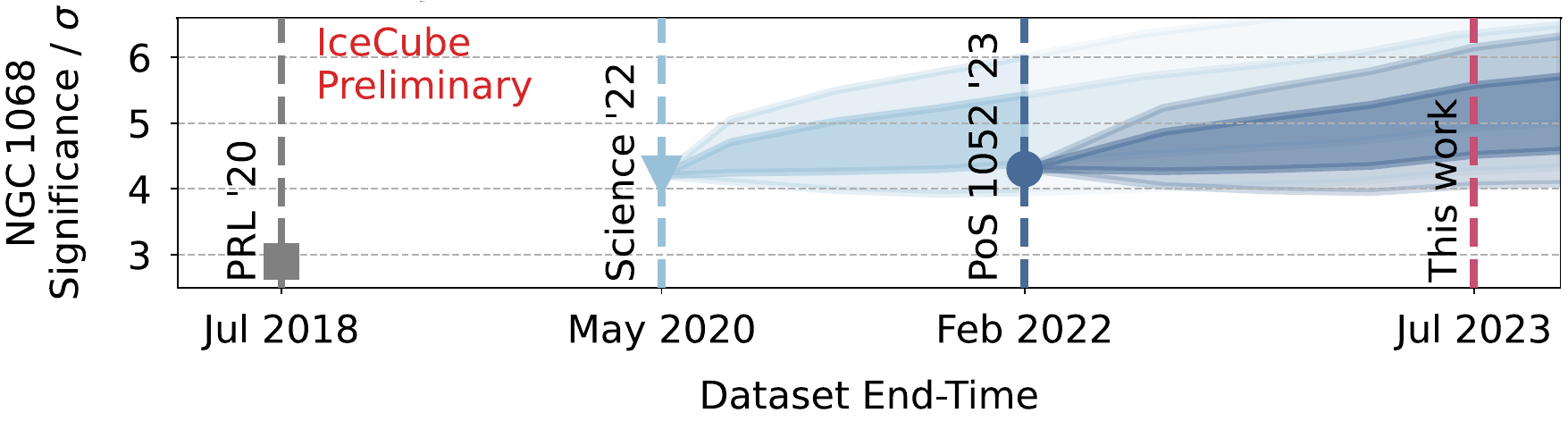}
    \caption{Evolution in the significance of IceCube observations of NGC~1068 as a function of time, including projections to future work assuming steady neutrino emission over time. Figure adapted from \cite{IceCube:2023_chiara}, reproduced with permission.}\label{Fig:NGC1068_significance_evolution}
\end{figure*}

The $4.2\,\sigma$ evidence of TeV neutrino emission from NGC 1068 (Section \ref{sec:neutrino}), combined with multi-wave\-length observations
(Sections \ref{sec:radio} - \ref{sec:gamma-ray}), especially in the X-ray and $\gamma$-ray bands, can potentially open up a new window on the HE and hadronic processes taking place in the innermost region(s) of AGN. Due to the limited IceCube angular resolution, this would not be possible with neutrinos alone. Several avenues complement each other in realizing this opportunity, some of which are being pursued already. As shown in Fig. \ref{Fig:NGC1068_significance_evolution}, the significance of the neutrino observations from NGC 1068 could reach the discovery threshold of $5\,\sigma$, commonly used in particle physics, in upcoming work \cite{IceCube:2023_chiara} aimed at extending 
the analysis 
 from nine to thirteen years.  According to preliminary projections, including refinements to the analysis methods, the significance is expected to fall within $4.5\,\sigma$ to $5.5\,\sigma$ ($68\%$ C.L.). 

The extent to which the processes that produce neutrinos in NGC~1068 generalize to the wider AGN population is presently unknown. It is therefore of the utmost importance to identify and study other neutrino-producing AGN. Assuming a connection with the plasma in AGN coronae, studies of Seyfert galaxies that host X-ray bright AGN (after accounting for X-ray absorption) have been encouraged \cite{Kheirandish:2021wkm}. 
Due to IceCube's markedly higher sensitivity to neutrino sources situated in the Northern Sky as opposed to those in the Southern Sky, the two regions are usually studied separately. Combining the model of refs. \cite{Murase2020, Kheirandish:2021wkm} with X-ray measurements reported by the BAT AGN Spectroscopic Survey (BASS) \cite{Ricci2017}, two corresponding IceCube searches are underway \cite{IceCube:2023_qinrui, IceCube:2023_shiqi}. Promising 
targets have been identified, and include NGC 4151 and NGC 4388 in the North, as well as the Circinus Galaxy in the South. Preliminary results indicate that the data associated with the selection of Seyfert galaxies in the Northern Sky, in particular NGC 4151 and CGCG 420-015, is inconsistent with the neutrino background at the $2.7\,\sigma$ level of significance, but not exactly as predicted by the combination of the chosen corona model \cite{Kheirandish:2021wkm} and the underlying X-ray measurements. For example, no potential signal was found for NGC 4388, thereby excluding the most optimistic model prediction of $\sim 20$ neutrinos at more than $90\%$ confidence level \cite{IceCube:2023_qinrui}. These results motivate further studies with ample opportunities for exciting discoveries but are currently insufficient to claim evidence for the tested scenario. 
A third IceCube analysis, looking for a correlation between IceCube neutrinos and hard X-ray AGN in general, reported NGC 4151 as the most interesting candidate source with a preliminary significance of $2.9\,\sigma$ \cite{IceCube:2023_sreetama}. For Compton-thick AGN, the estimated intrinsic X-ray luminosities come with large systematic uncertainties which influence directly the predicted neutrino fluxes. The estimation can be improved by including NuSTAR data for such objects  \cite{Tanimoto:2022wrq}, although in most cases it would still be rather model-dependent, since a direct view of the intrinsic luminosity is hampered even at very high energies. One example is NGC 3079, which may be brighter in neutrinos than estimated originally \cite{Neronov_2024}, adding to the growing indications of neutrino emission from AGN.

These on-going efforts would benefit from observations of these candidate sources in the MeV band by future facilities
\cite{deangelis2018,Caputo:2022xpx}, as discussed 
in Section \ref{sec:powers}, and reduced uncertainties in the distance estimates of nearby AGN.
Moreover, future neutrino detectors in the Northern hemisphere, such as GVD ({\url{https://baikalgvd.jinr.ru/}}), KM3NeT ({\url{https://www.km3net.org/}}), P-ONE ({\url{https://www.pacific-neutrino.org/}}), and TRIDENT ({\url{https://trident.sjtu.edu.cn/en}}), will be vital to provide stronger constraints on interesting targets in the Southern Sky.

\section{Summary and Conclusions}\label{sec:summary}

The detection of IceCube neutrinos from NGC 1068 was surprising because non-jetted AGN are typically dominated by thermal emission \cite{Padovani_2017}. Contrary to early ideas  \cite{berezinsky1977,eichler1979,silberberg1979}, recent work has primarily focused on jetted AGN, known for their predominantly non-thermal emission, as potential sources capable of accelerating protons to the energies required for neutrino production. 

NGC 1068 is a complex source with four distinct components: a SB, a jet, a molecular outflow, and an active BH. While each of these components could potentially be a source of neutrinos, a comprehensive analysis of the multi-messenger characteristics of this object led us to rule out the first three as possible neutrino emitters. This exclusion process ultimately points to the region near the BH as the sole environment where both proton acceleration and photon interactions can occur at the required intensity.

It is worth noting that the CRs possibly associated with NGC 1068 are of relatively medium energy
($\sim 40 - 400$ TeV), also as compared to the
much higher energies required for TXS 0506+056 ($\sim 800$ 
TeV -- 90 PeV) \cite{Padovani_2018}. This implies that non-jetted and jetted 
AGN might be complementary neutrino emitters, explaining at least part of the IceCube astrophysical background emission at the low- and 
high-energy end respectively. 

The general scenario emerging from the multi-messenger study reported in this paper can be then summarized as follows: 
1. the powerful inner region, possibly the AGN corona, is required to explain neutrino emission through photo-hadronic interactions; 2. this same region can also provide the photon-rich environment needed to absorb the neutrino-associated $\gamma$ rays, as these are not detected; 3. the outer region (SB, AGN wind) dominates the GeV $\gamma$-ray flux. 
It is important to keep in mind, however, that our knowledge of the X-ray corona is still limited, and this issue may not be entirely resolved.

We stress that the main conclusion we reach for NGC 1068,
namely that the neutrino emitter must be hidden in a $\gamma$-ray opaque environment, applies also to at least some
of the sources responsible for the diffuse background emission
observed by IceCube, as otherwise their $\gamma$-ray emission
would violate the {\it Fermi}-LAT constraints on the isotropic $\gamma$-ray background \cite{Murase2016}.

Future observations in the MeV energy band could probe the coronal activity, where the absorbed GeV - TeV $\gamma$ rays are expected to be reprocessed. Further studies with IceCube and other neutrino observatories will help constrain the connection between neutrinos and AGN and reveal the nature of the central engine powering the region surrounding the supermassive BH. In particular, the quest for the origin of the particles responsible for neutrino production and more specifically their spectrum, a crucial parameter to assess the energetics of the phenomena involved, would benefit from measurements of the neutrino flux down to energies $< 1$ TeV, which is however very challenging from the observational point of view. Such a measurement would provide a direct test of the acceleration mechanisms discussed above.

\section*{Data availability}
All data presented in this study are included in the article or are freely available.

\section*{Acknowledgments}

The idea for this review came during a Topical
Workshop on NGC 1068 organised by ER, CB, and PP and held at the 
Munich Institute for Astro-, Particle and BioPhysics in Garching on March 6 - 10, 2023. The authors, as organisers and invited speakers, wish to thank all the participants to the meeting for the stimulating atmosphere and fruitful discussions.

This work is supported by the Deutsche Forschungsgemeinschaft (DFG, German Research Foundation) through grant SFB 1258 ``Neutrinos and Dark Matter in Astro- and Particle Physics'' and by the Excellence Cluster ORIGINS which is funded by the  DFG under Germany's Excellence Strategy - EXC 2094. 
BE acknowledges support from the DFG within the Collaborative Research Center SFB 1491 ``Cosmic Interacting Matters - From Source to Signal''.
EP acknowledges support from the Villum Fonden (No.~18994) and from the European Union's Horizon 2020 research and innovation program under the Marie Sklodowska-Curie grant agreement No. 847523 'INTERACTIONS'.
SG and EP were also supported by Agence Nationale de la Recherche (grant ANR-21-CE31-0028).

\section*{Author contributions}

PP coordinated the work, contributing mostly to Sections 
\ref{sec:intro} -- \ref{sec:powers}, \ref{sec:summary}, and 
Methods \ref{app:facts} and \ref{app:radio_supp}}, 
and collaborated with ER on finalizing the paper. ER also contributed to Sections \ref{sec:intro} and \ref{sec:summary}. KH worked on Section \ref{sec:sub-mm}, while VGR and TS focused on Section \ref{sec:NIR}, SB on Section \ref{sec:X-ray}, MA and AL on Section \ref{sec:gamma-ray}, CB on Sections \ref{sec:neutrino} and \ref{sec:outlook}, SG on Section \ref{sec:powers}, EP, BE, DG, and AL on Section \ref{sec:theor_models}, PB on Section \ref{sec:theor_models} and Methods  \ref{app:accel},
and HN on Section \ref{sec:outlook}. All authors participated in discussions regarding the manuscript.

\section*{Competing interests}

The authors declare no competing interests. 

\section*{Additional information}

Correspondence should be addressed to Paolo Padovani ~$<$ppadovan(at)eso.org$>$.

\begin{appendices}

\section*{Methods}
    
\section{The distance to NGC 1068}\label{app:facts}

There seems to be some 
confusion in the literature as regards the distance of NGC 1068, 
a fundamental parameter, which 
cannot be simply derived from its redshift ($z=0.00379$) since it is 
relatively nearby and therefore subject to the gravitational 
pull of local overdensities. These 
produce so-called ``peculiar velocities'' in addition to, 
and typically larger than, the velocity due to cosmic expansion. 
Most papers use a luminosity distance $D_{\rm L} = 14.4$ Mpc, which 
can be traced back to a 1997 paper \cite{Bland-Hawthorn_1997}, which 
refers back to a 1988 paper by Brent Tully \cite{Tully_1988}. But a 2008 paper led by the same author instead provides a 
distance $D_{\rm L} = 10.1 \pm 1.8$ Mpc \cite{Tully_2008}, which 
is consistent with the very recent value of $11.1 \pm 0.5$ 
Mpc based on a different and independent method \cite{Tikhonov_2021}. Hence, 
current evidence supports a relatively low distance and excludes a value as 
large as 14.4 Mpc (we thank Dmitry Makarov from HyperLeda 
(\url{https://leda.univ-lyon1.fr/}) for his help on this issue). 
In this paper, therefore, we use $D_{\rm L} = 10.1$ Mpc. 
We have converted all powers in the literature using  
this value, that is, we have reduced by 50\% values derived using 
$D_{\rm L} = 14.4$ Mpc and by a factor of 2.6 those estimated assuming 
NGC 1068 is in the Hubble flow [$D_{\rm L} = 16.3$ Mpc for a standard $\Lambda$CDM 
cosmology (Hubble constant $H_0 = 70$ km s$^{-1}$ Mpc$^{-1}$, matter
density $\Omega_{\rm m,0} = 0.3$, and dark energy density $\Omega_{\Lambda,0} 
= 0.7$.)]. 

\section{Radio decomposition and jet power}\label{app:radio_supp}

The relative contributions of the SB and jet 
components can be estimated through the so-called 
radio--FIR correlation, that is, the very tight relation between 
radio power at 1.4 GHz, $L_{\rm 1.4 GHz}$, and the $8 - 1,000~\mu m$ luminosity, 
$L_{\rm FIR}$, which is valid for star-forming (SF) galaxies \cite{Kennicutt_1998}. 
Using $\log L_{\rm FIR} = 11$ (in units of solar bolometric luminosity 
$L_{\rm \odot} = 3.83 \times 10^{33}$ erg s$^{-1}$)  \cite{Sanders_2003}, 
and $\log L_{\rm 1.4 GHz} = \log L_{\rm FIR} + 11.47$ \cite{Bonzini_2015}, 
we derive $L_{\rm SB, 1.4 GHz} \sim 3.0 \times 10^{22}$ W Hz$^{-1}$, 
which implies a roughly equal share between the jet and the SB region at 1.4 GHz. 
The total star formation rate (SFR) derived in two different
and independent ways from $L_{\rm FIR}$ and $L_{\rm SB, 1.4 GHz}$ 
using eqs. (2) and (3) of ref. \cite{Bonzini_2015} ($\sim 10.3$ and $17.4~M_{\odot}$ yr$^{-1}$ respectively), puts NGC 1068 in the SB regime 
(ref. \cite{Leroy_2021} finds an even higher SFR $\sim 22.8~M_{\odot}$ 
yr$^{-1}$ using UV and near-IR [NIR] data). 
We can also estimate the {\it total} jet power from $L_{\rm jet,1.4 GHz}$. This is done  
by studying the X-ray cavities which surround many massive galaxies and
provide a direct measurement of the mechanical energy released by AGN through 
work done on the hot, gaseous halos. We use eq. (1) from ref. \cite{Cavagnolo_2010}
and derive $L_{\rm jet} = 10^{42.9\pm1.0}$ erg s$^{-1}$ from $L_{\rm 
jet,1.4 GHz}$. Using the more recent ref. \cite{Heckman_2023} we get $L_{\rm 
jet} = 10^{42.2}$ erg s$^{-1}$, which is consistent with our result within 
its (large) uncertainty.

\section{Stochastic acceleration in the corona}\label{app:accel}

Second order Fermi acceleration in the presence of turbulence can be described with an arbitrarily high level of complication. A modern approach to this process, based on first principles, has recently been put forward in ref. \cite{Lemoine_2022}. Here we adopt a simplified treatment that provides us with some interesting insights. We assume that turbulence in the corona is described by a power spectrum per unit logarithmic wavenumber $k$ as ${\cal F}(k)=(q-1)(k/k_0)^{1-q}$, where $q=3/2$ ($q=5/3$) for Kraichnan (Kolmogorov) turbulence and $k_0^{-1}=1/R_c$ is the scale where the turbulence is assumed to be injected. We also assume that the magnetic field in the corona is fully turbulent and $\beta_P\sim 1$. The motion of charged particles in this turbulence is diffusive with a pitch angle diffusion coefficient $D_{\mu\mu}\approx \Omega {\cal F}(k)$, with $k=1/r_L(p)$ and $\Omega$ is the gyration frequency. 

The spatial diffusion of non-thermal particles also leads to diffusion in momentum space, the standard second order Fermi acceleration. The diffusion coefficient in p-space is $D_{pp}\approx (v/v_A)^2 p^2 D_{\mu\mu}$, which corresponds to an acceleration time:
\begin{equation}
    \tau_{acc}=\frac{p^2}{D_{pp}}= 
0.4 \rm s \left( \frac{p}{GeV/c}\right)^{1/2} M_7^{3/4} \tau_T^{1/4} (\eta/30)^{3/2},
\end{equation}
where we specialized the estimate to a Kraichnan spectrum, $q=3/2$, and relativistic particles. 

As discussed by ref. \cite{Murase2020}, particle acceleration in a hidden source like NGC 1068 is limited by Bethe-Heitler pair production, being slightly faster than $p - p$ and $p - \gamma$ interactions. These processes limit the maximum energy to $E_{max}\lesssim 100$ TeV (though dependent upon the intrinsic luminosity of the source). 
For $E\ll E_{max}$, one could assume that acceleration occurs in a stationary regime and neglect the effect of losses, thereby allowing us to infer the spectrum of accelerated particles. It is easy to see that since $D_{pp}\propto p^q$, the solution of the diffusion equation in momentum space must also have a power law shape, $n(p)\propto p^{-\beta}$ with $\beta=q+1$, namely, for relativistic particles, $N(E)\propto p^2 n(p)\sim E^{1-q}$. 
For typical spectra of turbulence, one can see that the spectra of accelerated particles are much harder than those expected from shock acceleration, with a cutoff dictated by energy losses. In other words, second order acceleration pushes all energy toward $E_{max}$, thereby creating a spectrum that satisfies the energetic conditions discussed above, producing at the same time a steep neutrino spectrum close to the cutoff.

\end{appendices}

\bibliography{NGC1068_review_final}

\end{document}